\documentclass[hyphens,sigconf,10pt,nonacm]{acmart}
\usepackage{booktabs}

\usepackage{amsmath}
\renewcommand{\cite}[1]{\errmessage{Don't use cite, you probably want citep}}
\usepackage{verbatim}
\usepackage{xspace}
\usepackage{balance}
\setkeys{Gin}{keepaspectratio=true,clip=true,draft=false,width=\linewidth}
\graphicspath{{./imgs/}}
\newcommand{\Hi}{\textsc{Hi}\xspace}
\newcommand{\Lo}{\textsc{Lo}\xspace}

\begin{document}
  \sloppy

  \renewcommand{\sectionautorefname}{Sect.}
  \renewcommand{\subsectionautorefname}{Sect.}
  \renewcommand{\subsubsectionautorefname}{Sect.}
  \renewcommand{\appendixautorefname}{Appendix}
  \renewcommand{\Hfootnoteautorefname}{Footnote}

  \acmConference{}{}{}

\title{Can We Prove Time Protection?}
  \author{Gernot Heiser}
  \orcid{0000-0002-7069-0831}
  \affiliation{\institution{UNSW Sydney} and \institution{Data61, CSIRO}}
  \email{gernot@unsw.edu.au}
  \author{Gerwin Klein}
  \affiliation{\institution{Data61, CSIRO} and \institution{UNSW Sydney}}
  \email{gerwin.klein@data61.csiro.au}
  \author{Toby Murray}
  \affiliation{\institution{University of Melbourne} and \institution{Data61, CSIRO}}
  \email{toby.murray@unimelb.edu.au}

  \begin{abstract}
    Timing channels are a significant and growing security threat in
    computer systems, with no established solution. We have recently
    argued that the OS must provide \emph{time protection}, in analogy
    to the established memory protection, to protect applications from
    information leakage through timing channels. Based on a
    recently-proposed implementation of time protection in the seL4
    microkernel, we investigate how such an implementation could be
    formally proved to prevent timing channels. We postulate that this
    should be possible by reasoning about a highly abstracted
    representation of the shared hardware resources that cause timing channels.
  \end{abstract}
  \maketitle

\section{Introduction}\label{s:intro}

Timing channels are a major threat to information security, they exist
where the timing of a sequence of observable events depends on secret
information~\citep{Wray_91}. The observation might be of an externally
visible event, such as the response time of a server, and might be
exploitable over intercontinental distances \citep{Cock_GMH_14}. Or it might only be
locally observable, i.e.\ by a process or VM co-located on the same
physical machine, which still enables remote attacks, if the observing
process has access to the network and is controlled by a remote
agent. The seriousness of the threat was recently highlighted by the
Spectre attacks~\citep{Kocher_HFGGHHLMPSY_19}, where speculatively
executed gadgets leak information via a covert timing channel.

The secret-dependence of events may have \emph{algorithmic} causes, e.g.\
crypto implementations with secret-dependent code paths. Or they may
result from interference resulting from competing access to limited
hardware resources, such as caches; there exists a wide variety of such
\emph{micro-architectural channels}~\citep{Ge_YCH_18}.

Whether algorithmic or micro-architectural, those channels represent
information flow across protection boundaries, i.e.\ \emph{the boundaries
are leaky}. Ensuring the security of these boundaries should be the job
of the operating system (OS); however, no contemporary, general-purpose
OS seems to be capable of if. Clearly, this is not an acceptable
situation, and we have recently called for OSes to provide \emph{time
  protection}~\citep{Ge_YH_18} as the temporal equivalent of the
well-established concept of memory protection.

Memory protection is a solved problem: the formal verification of seL4
proved, among others, that the kernel is able to enforce spatial
integrity, availability and confidentiality
\citep{Klein_AEMSKH_14}. This categorically rules out
information leakage via \emph{storage channels} (provided that the kernel is
aware of the state that can be used for such channels).  However, the
approach taken in the seL4 verification has no concept of time, and
therefore cannot make any claims about \emph{timing channels}.

\textbf{Our aim is  to rule out timing-channel leakage just as
  categorically as information flow via storage.} Put differently, \emph{we
aim to formally prove correct implementation of time protection}. This paper investigates the
feasibility of, and prerequisites for, achieving the stated aim. Obviously we
would not bother writing this paper if we were not convinced that it
is feasible to achieve our aim, under certain conditions, which come
down to hardware satisfying certain requirements.  We have recently
demonstrated that not all recent processors satisfy these requirements, resulting in a call for a new, security-oriented
hardware-software contract~\citep{Ge_YH_18}. We claim that, for hardware that honours
this contract, we will be able to achieve our aim of proving time
protection, and thus eliminate micro-architectural timing channels.

Note that other physical channels, such as power draw, temperature, or
acoustic or electromagnetic emanation, are outside the scope of this work. 

\section{Threat Scenario}

The basic problem we are concerned with is a secret held by one
security domain, \Hi, being leaked to another domain, \Lo,
which is not supposed to know it. The leaking might be intentional, by
a bad actor (Trojan) inside \Hi, constituting a \emph{covert channel}. Or
it can be unintentional, via a \emph{side channel}. Note that \Hi, \Lo
are relative to a
particular secret, we do not assume a hierarchical security policy
such as \citet{Bell_LaPadula_76}, and there may be other
secrets for which the roles of the domains are reversed. It is the
duty of the OS to prevent any unauthorised information flow, no matter
what the system's specific security policy might be.

Our notion of a security domain refers to a subset of the system
which is treated as an opaque unit by the system's security policy
(i.e.\ intra-domain information flow is not restricted by the
policy). In OS terms, a domain consists of one or more (cooperating) processes.

We assume that the OS provides strong, verified memory protection, and
is free of storage channels, seL4 being an example. Our primary
concern is micro-architectural channels, i.e.\ channels that exploit
competition for finite hardware resources that are abstracted away by
the instruction-set architecture (ISA), the classic hardware-software
contract. This means that algorithmic channels are not our primary
concern, but we will discuss in \autoref{s:algorithmic} how time protection can be employed
to remove such channels (within limits).

Like memory protection, time protection is a black-box OS mechanism,
that provides \emph{mandatory security enforcement} without
relying on application cooperation.

For realism, i.e.\ to ensure that contemporary hardware is at least
close to satisfying the requirements of time protection (and can fully
satisfy them with minor enhancements) we limit our scope in one
important way: we do not (yet) attempt to prevent \emph{covert channels}
through stateless interconnects. Such channels, exploiting the finite
bandwidth of interconnects through concurrent competing access,
are trivial to implement: a Trojan running on one core signals by
modulating its use of interconnect bandwidth, and a spy running on a
different core measures the remaining bandwidth by trying to saturate
the shared interconnect. Such channels can only be prevented with hardware
support that is not available on any contemporary mainstream
hardware.\footnote{Intel recently introduced 
\emph{memory bandwidth allocation} (MBA) technology, which  imposes
\emph{approximate} limits on the memory bandwidth available to a
core~\citep{Intel_64_IA-32:asdm2_325383}. While this represents a step towards
bandwidth partitioning, the approximate enforcement is not
sufficient for preventing covert channels.} We will be able to extend time
protection
in a fairly straightforward way, should such hardware support (or at least an accepted
model for it) become available.

An obvious example of the excluded scenario would be a covert channel between
two virtual machines (VMs) concurrently executing on different cores
of the same processor on a public cloud. Such a covert channel is not
a particular concern, as the Trojan in the victim VM does not need the
co-located spy, as it can communicate by other means, e.g.\ modulating
its network communication. Side channels are a real concern in the
cloud scenario, but stateless interconnects reveal no address
information. As a consequence, no such side channels have been
demonstrated to date~\citep{Ge_YCH_18}, and they are likely
impossible.

\section{Timing-Channel Mechanisms}

There are two ways in which \Lo may learn \Hi's secret: by timing
observable actions of \Hi, or by \Lo observing how its own execution
speed is influenced by \Hi's execution.

\subsection{Timing own progress}

This channel utilises the performance impact of interference between
processes resulting from competition for shared hardware resources,
especially stateful resources such as caches, TLBs, branch predictors
and pre-fetcher state machines. For example, \Lo's rate of progress
(performance) is affected by cache misses. If \Lo shares a cache with
\Hi (either time-sharing a core-private cache or concurrently sharing
a cache with \Hi's core), then the miss rate will depend on \Hi's
cache usage. If the cache is set-associative (which almost all of them
are nowadays), then the pattern of cache misses will also reveal
address information from \Hi. Such address information supports the
implementation of side channels with potentially high bandwidth, e.g.\
where the secret is used to index a table~\citep{Ge_YCH_18}.

An effective exploitation of such a channel is the
\emph{prime-and-probe} technique~\citep{Percival_05,  Osvik_ST_06}. Here \Lo fills
the cache by traversing a buffer large enough to cover the cache
(prime phase). After or while \Hi is executing, \Lo traverses the
buffer again, monitoring the time taken for each access (probe phase);
a long latency indicates a conflict miss with \Hi's cache
footprint. The address of the missing access reveals the index bits of
\Hi's access.

Prime-and-probe can be used as a high-bandwidth covert channel, where
\Hi explicitly encodes information into the memory addresses accessed,
or as a side channel, where the encoding is implicit in \Hi's normal
execution (e.g.\ via a secret-derived array index). It can be used for
time-shared (core-private) caches as well as caches shared between
cores.

\subsection{Timing \Hi events}

\begin{figure}[t]
  \centering
  \includegraphics[width=\linewidth]{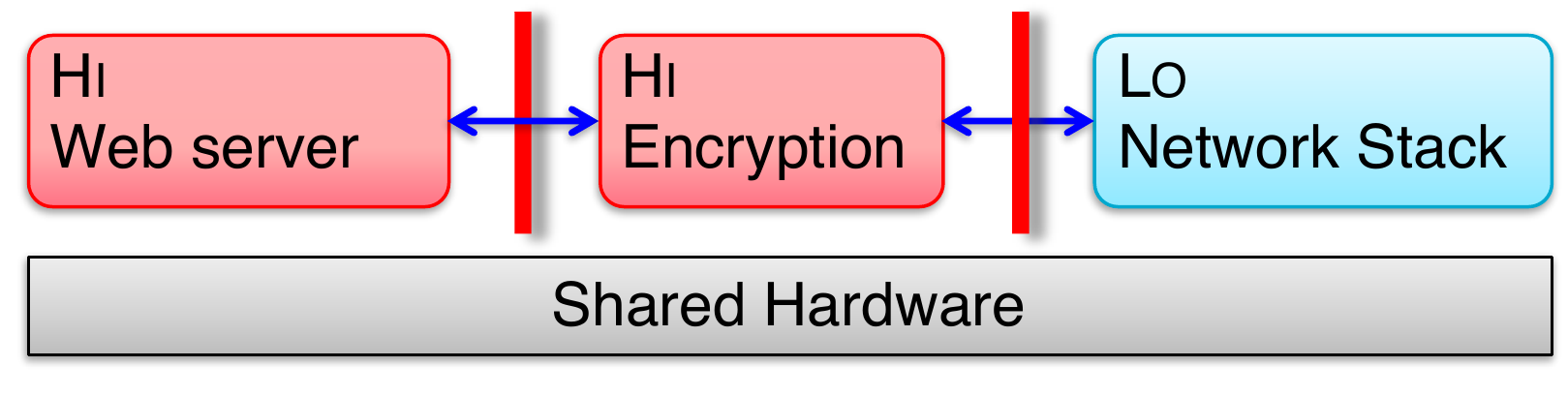}
  \caption{Encryption engine as a downgrader.}
  \label{f:downgrader}
\end{figure}

On a first glance this might seem like a silly case, why worry about a
covert channel if there is an overt one, such as message passing?
However, this situation is in fact common: \Hi might be a
\emph{downgrader}, an entity trusted to handle secrets and decide
which can be safely declassified. A common example is a crypto
component, which encrypts secrets, e.g.\ from a web server, and
publishes the encrypted text, by handing them to a network unit; this
is shown in \autoref{f:downgrader}.

In this case, the leakage might be resulting from an algorithmic
channel (e.g.\ a crypto implementation with secret-dependent
execution), a Trojan modulating the speed of the encryption 
(possibly via micro-architectural interference), or the server itself
leaking through the timing of messages to the crypto component.

Time protection here must make execution time
deterministic, meaning that message passing or context switching
happen at pre-determined times. Obviously, the OS can only provide the
mechanism here (deterministic switch/delivery time), not the policy
(the time of the switch). This must be set by the system
designer or security officer, taking into account issues like the
worst-case execution time (WCET) of the encryption.

\citet{Cock_GMH_14} have proposed a possible model: a synchronous IPC
channel switches to the receiver only once the sender domain has
executed for a pre-determined minimum amount of time. It is then left
to the system designer to determine a safe time threshold.

\section{Defences and Limitations}

\subsection{Flushing and partitioning hardware}\label{s:aisa}

As micro-architectural timing channels result from competition for
(non-architected) hardware resources, eliminating them requires
removing the competition. This means the OS must either partition
those resources between security domains, or reset them to a defined,
history-independent state between accesses from different domains.

Resetting, e.g.\ flushing caches, only helps where accesses from
different domains are
separated in time, i.e.\ for time-shared resources. In other
words, resetting only works for resources that are private to an
execution stream. In the absence of hyperthreading, this applies to
core-local resources, such as the L1 caches, private L2 caches (on
Intel hardware), TLBs, branch predictors, and core-local prefetchers.

Partitioning is the only option where concurrent accesses happen,
i.e.\ for caches shared between cores. It would also be the only
option for core-local state when hyperthreading is enabled. However,
no mainstream hardware supports partitioning of hardware resources
between hyperthreads, and such partitioning would seem fundamentally
at odds with the concept of hyperthreading, which is based on
improving hardware utilisation by sharing. Consequently there are a
plethora of side-channel attacks between
hyperthreads~\citep{Ge_YCH_18}. We have to conclude that
\emph{hyperthreading is fundamentally insecure}, and multiple hardware
threads must never be allocated to different security domains
(multi-threading a single domain is not a security issue).

Partitioning of shared (physically-addressed) caches is possible
without extra hardware support by using page
colouring~\citep{Kessler_Hill_92, Lynch_BF_92, Liedtke_HH_97}. This
uses the fact that the associative lookup of a large cache forces a
page into a specific subset of the cache, so only pages mapping to the
same subset, said to have the same colour, can compete for cache
space. By ensuring that different security domains are allocated
physical page frames of disjoint colours, the OS can partition the
cache between domains. Modern last-level caches have at least 64
different colours.

In general, \emph{micro-architectural timing channels can be prevented if
all shared hardware can be either partitioned of flushed by the
OS}, with flushing the only option where accesses are
concurrent. Together with a few other conditions
outlined by \citet{Ge_YH_18}, these form part of a security-oriented
hardware-software contract, called the aISA (augmented ISA), that allows the OS to prevent timing
channels. The ISA alone is an insufficient contract for ensuring
security~\citep{Heiser_18, Hill:ISA-blog}.

\subsection{Implementing time protection}\label{s:impl}

We have recently proposed an implementation of time protection in
seL4, for hardware that conforms to a security-oriented
aISA~\citep{Ge_YCH_19}. It uses cache colouring to partition shared
caches. As even read-only sharing of code is sufficient for creating a
channel~\citep{Gullasch_BK_11, Yarom_Falkner_14}, we also colour the
kernel image. This is achieved by a policy-free \emph{kernel clone}
mechanism, which allows setting up a domain-private kernel image in
coloured memory.

We flush time-shared micro-architectural state on each domain switch
(but not on intra-domain context-switches). For writable
micro-architectural state (e.g.\ the L1 data cache), the latency of
the flush is itself dependent on execution history (number of dirty
lines), which would create a channel. We avoid this channel by padding
the domain-switch latency to a fixed value. For generality (see
\autoref{s:algorithmic}) we make determining the padding time not the
job of the OS, but an attribute of the switched-from security domain,
controlled by the system designer. Specifically, we specify that the
next domain will not start executing earlier than the previous
domain's time slice plus the padding time.

The padding time should obviously be at least the worst-case latency
of the flush, but also needs to account for any delay of the handling of the
preemption-timer interrupt by other kernel entries (resulting from
system calls or interrupts).

Finally, interrupts could also be used as a channel, if the Trojan
triggers an I/O such that its completion interrupt fires during \Lo's
execution~\citep{Ge_YCH_19}.  We prevent this by partitioning
interrupts (other than the preemption timer) between domains, and keep
all interrupts masked that are not associated with the
presently-executing domain.

\subsection{Preventing algorithmic channels}\label{s:algorithmic}

Padding is a general mechanism that can also be used to prevent
algorithmic channels. In the scenario of \autoref{f:downgrader}, we
can pad the execution time of the downgrader to a safe value (an upper
bound of its execution time). In practice, this is very wastive if
padding is done by busy looping. To make it practical, another \Hi
process should be scheduled for padding. Obviously, that interim
process must be preempted early enough to allow the kernel to switch
domains without \emph{exceeding} the pad time (as this might introduce
new channels). This may not be straightforward to
implement, but it is clearly possible.

\section{Proving Time Protection}

At first glance, one might expect that proving time protection is a
hopeless exercise. After all, the precise interaction between
microarchitectural state and execution latency is unspecified for
modern hardware platforms, and the latency of some instructions may
vary by orders of magnitude depending on hardware state. Formally reasoning about precise execution
latencies is therefore infeasible~\citep{Klein_MGSW_11}.

However, we argue that reasoning about the exact latency of executions is
unnecessary. \textbf{The key insight  is that these channels are effected
by shared hardware resources, and if we can prove that no sharing
happens, there can be no timing channels.}
Consequently, proving temporal isolation requires formal models of
microarchitectural state, but these can be kept abstract,
providing only detail to identify
resources that need to be partitioned (and how such partitioning is
performed), and state that must be reset (and how to reset it). That is,
we do not need to know how long an instruction will take to execute, only
which micro-architectural state its execution time depends on and how this  state behaves wrt.\ partitioning and flushing.

For partitionable state, temporal isolation becomes a
\emph{functional} property (namely an invariant about correct partitioning)
that can be verified without any
reference to time, meaning existing verification techniques apply.

For state that requires flushing, correct application of the flush is also a
functional property.  As mentioned in \autoref{s:impl},
the latency of flushing operations
themselves needs to be hidden by the OS, by padding its execution.
Correct padding can be verified with a relatively simple
formalisation of hardware clocks, which allows verifying padding
time by simply comparing time stamps, reducing this to a
functional property as well.

Once timing-channel reasoning is reduced to the
verification of functional properties, it should be possible to
integrate it into existing proof frameworks of storage-channel freedom,
such as seL4's information flow proofs~\citep{Murray_MBGBSLGK_13}.

Indeed, under this approach \textbf{timing-channel reasoning
is transmuted into reasoning about storage channels}, reducing it to
a solved problem, and also enabling
reasoning about timing-channels without reference to precise execution time.
This possibility
may seem surprising, but it is known that the distinction between
storage and timing channels is not fundamental, but refers to the mechanisms used
for exploitation~\citep{Wray_91}. In our case we transform the
temporal interference problem into a spatial one, by reasoning about
the shared hardware resources which the channels exploit.

\subsection{Hardware formalisation}

Carrying out these proofs requires a model of the shared hardware
resources (the \emph{microarchitectural model})
that influence execution latencies,
as well as a simple model of a hardware clock (the \emph{time model})
to allow reasoning about
elapsed time intervals. Naturally these models are interrelated: how much
an execution step advances the hardware clock naturally depends on the
microarchitectural state that influences execution time.

Crucially, a precise description of this interaction is not
necessary. Instead, the interaction can be faithfully yet feasibly
modelled as follows. Firstly, the microarchitectural model must delineate
the partitionable state from the flushable state, and all microarchitectural
state must be partitionable or flushable (\autoref{s:aisa}). Secondly, the time model, which
captures how far time advances on each
execution step, is defined as a \emph{deterministic yet unspecified}
function of the microarchitectural state. Then, when the microarchitectural
state is properly partitioned and flushed, one can prove that a security
domain's execution time cannot be influenced by other domains
(see \autoref{sec:proof} below).

This construction neatly reflects the basic assumptions that (i) the hardware
provides sufficient mechanisms to partition and flush microarchitectural
state between security domains, that (ii) such mechanisms work correctly, and
that (iii) these account for all microarchitectural state that influences
execution time.

\subsection{Information-flow proofs}
\label{sec:proof}

With these models in hand, time protection can then be proved by
showing that there is no way in which the execution of one domain
can affect the execution timing of another domain.

Specifically
the proofs must show that all resource partitioning and flushing is
applied at all times and not bypassable, and that domain-switches
(flushing) is correctly padded to a constant amount of time (under the
assumption that the padding value, obtained by a separate analysis, is sufficient). These
proofs can then be integrated with existing storage-channel freedom proofs
to derive the absence of
timing channels as follows. 

Without loss of generality, fix some domain (\Lo) and consider one of its
execution steps for which we show its timing cannot be influenced by
another domain (\Hi). There are two possibilities:
(Case 1) either it is an ordinary user-mode instruction, or (Case 2)
it is a trap (a system call, exception, or interrupt arrival).
For Case 1, the execution time given by the time model
will be affected by the shared hardware
resources in the microarchitectural model. Recall that this effect can be
modelled by an unspecified deterministic function from the state of
the microarchitectural model to an elapsed (symbolic) time value. For an individual instruction
this function will examine the state of the instruction cache, namely the
cache set identified by the program counter, and the state of the data cache
for any memory address accessed
by that instruction. Since the access does not fault (otherwise it would be a trap), all such memory accesses
must lie within the physical memory of the current domain and thus
within areas of the cache that cannot be affected by other partitions
(due to correct cache partitioning by the kernel, or correct flushing,
e.g.~for the on-core L1 cache).

A similar argument applies to other microarchitectural resources.
Thus the resulting execution time cannot be affected by other
partitions.

For Case 2, we distinguish two sub-cases: The trap is either (Case 2a)  a
system call or exception, or (Case 2b) it is the arrival of a timer interrupt
signalling a switch to the next domain.
For Case 2a,
the execution time depends on the state of the instruction cache
wrt.\ the kernel instructions executed, plus the data cache for any
data accessed. However, in a partitioned system with the kernel correctly
cloned as in \autoref{s:impl}, 
the former cannot be affected by other partitions and the latter
accesses only data of the current domain. The only remaining state
that might be accessed is global kernel data, which we will
prove is accessed deterministically and whose cache state
after a domain switch is independent of prior \Hi activity (due to
correct flushing). 
Thus a similar, if naturally more involved, argument applies
as to the user mode case (Case 1). Incorporating general (i.e.\ non-timer)
interrupts we believe is also possible, by partitioning the interrupt set
as covered in \autoref{s:impl}.
For Case 2b, we invoke the proof of the constant-time domain switch
property.  $\square$

\smallskip

Note that by reflecting elapsed time as a value in the state of the time model
(updated by an unspecified function of the microarchitectural model),
timing-channel reasoning is reduced to storage-channel reasoning, and indeed
time protection itself can be phrased and proved
akin to storage-channel freedom via
a suitable noninterference property~\citep{Murray_MBGK_12}.

\subsection{TLB}

The TLB is an example where the principles of partitioning and flushing can
already be observed in a formal model for pure functional correctness: while
not yet suitable for reasoning about timing, \citet{Syeda_Klein_18} provide a
logic for functional correctness under an ARM-style TLB. For instance, it is
easy to show in this model that page tables modifications under one address
space identified (ASID) do not affect TLB consistency for any other ASID.
This is the kind of partitioning theorem we would make use of for timing-relevant state.

The model in this work is a high-level abstraction of the TLB
proved sound with respect to a low-level model that would be
infeasible to reason about directly. We propose the same for timing behaviour.
Instead of reasoning about a detailed low-level architecture model with precise
timing information, we only record the information needed for
timing-independence.

\section{Conclusions}

We conclude that proving time protection should be possible with
established formal methods, thanks to the key insight that they result
from spatial-type micro-architectural resources, and can thus be
treated as storage channels. This requires some reasoning about those
hardware resources, but we expect to get away with very high-level
abstractions. The key challenge is to achieve agreement on a
hardware-software contract that makes it at least possible to remove
timing channels. We are clearly at the mercy of processor
manufacturers here!

\balance
{\sloppy
  \bibliographystyle{ACM-Reference-Format}
  \bibliography{references}


\providecommand{\noopsort}[1]{}\providecommand{\url}{\error{The bib files now
  require the `url' package!}}\providecommand{\NoRemove}{}
\begin{thebibliography}{22}


\ifx \showCODEN    \undefined \def \showCODEN     #1{\unskip}     \fi
\ifx \showDOI      \undefined \def \showDOI       #1{#1}\fi
\ifx \showISBNx    \undefined \def \showISBNx     #1{\unskip}     \fi
\ifx \showISBNxiii \undefined \def \showISBNxiii  #1{\unskip}     \fi
\ifx \showISSN     \undefined \def \showISSN      #1{\unskip}     \fi
\ifx \showLCCN     \undefined \def \showLCCN      #1{\unskip}     \fi
\ifx \shownote     \undefined \def \shownote      #1{#1}          \fi
\ifx \showarticletitle \undefined \def \showarticletitle #1{#1}   \fi
\ifx \showURL      \undefined \def \showURL       {\relax}        \fi
\providecommand\bibfield[2]{#2}
\providecommand\bibinfo[2]{#2}
\providecommand\natexlab[1]{#1}
\providecommand\showeprint[2][]{arXiv:#2}

\bibitem[\protect\citeauthoryear{Bell and LaPadula}{Bell and LaPadula}{1976}]%
        {Bell_LaPadula_76}
\bibfield{author}{\bibinfo{person}{D.E. Bell} {and} \bibinfo{person}{L.J.
  LaPadula}.} \bibinfo{year}{1976}\natexlab{}.
\newblock \bibinfo{booktitle}{\emph{Secure Computer System: Unified Exposition
  and {Multics} Interpretation}}.
\newblock \bibinfo{type}{{T}echnical {R}eport} MTR-2997.
  \bibinfo{institution}{{MITRE} Corp.}
\newblock


\bibitem[\protect\citeauthoryear{Cock, Ge, Murray, and Heiser}{Cock
  et~al\mbox{.}}{2014}]%
        {Cock_GMH_14}
\bibfield{author}{\bibinfo{person}{David Cock}, \bibinfo{person}{Qian Ge},
  \bibinfo{person}{Toby Murray}, {and} \bibinfo{person}{Gernot Heiser}.}
  \bibinfo{year}{2014}\natexlab{}.
\newblock \showarticletitle{The Last Mile: An Empirical Study of Some Timing
  Channels on {seL4}}. In \bibinfo{booktitle}{\emph{ACM Conference on Computer
  and Communications Security}}. \bibinfo{address}{Scottsdale, AZ, USA},
  \bibinfo{pages}{570--581}.
\newblock


\bibitem[\protect\citeauthoryear{Ge, Yarom, Chothia, and Heiser}{Ge
  et~al\mbox{.}}{2019}]%
        {Ge_YCH_19}
\bibfield{author}{\bibinfo{person}{Qian Ge}, \bibinfo{person}{Yuval Yarom},
  \bibinfo{person}{Tom Chothia}, {and} \bibinfo{person}{Gernot Heiser}.}
  \bibinfo{year}{2019}\natexlab{}.
\newblock \showarticletitle{Time Protection: the Missing {OS} Abstraction}. In
  \bibinfo{booktitle}{\emph{Eurosys19}}. \bibinfo{publisher}{ACM},
  \bibinfo{address}{Dresden, Germany}.
\newblock


\bibitem[\protect\citeauthoryear{Ge, Yarom, Cock, and Heiser}{Ge
  et~al\mbox{.}}{2018b}]%
        {Ge_YCH_18}
\bibfield{author}{\bibinfo{person}{Qian Ge}, \bibinfo{person}{Yuval Yarom},
  \bibinfo{person}{David Cock}, {and} \bibinfo{person}{Gernot Heiser}.}
  \bibinfo{year}{2018}\natexlab{b}.
\newblock \showarticletitle{{A} Survey of Microarchitectural Timing Attacks and
  Countermeasures on Contemporary Hardware}.
\newblock \bibinfo{journal}{\emph{Journal of Cryptographic Engineering}}
  \bibinfo{volume}{8} (\bibinfo{date}{April} \bibinfo{year}{2018}),
  \bibinfo{pages}{1--27}.
\newblock


\bibitem[\protect\citeauthoryear{Ge, Yarom, and Heiser}{Ge
  et~al\mbox{.}}{2018a}]%
        {Ge_YH_18}
\bibfield{author}{\bibinfo{person}{Qian Ge}, \bibinfo{person}{Yuval Yarom},
  {and} \bibinfo{person}{Gernot Heiser}.} \bibinfo{year}{2018}\natexlab{a}.
\newblock \showarticletitle{No Security Without Time Protection: We Need a New
  Hardware-Software Contract}. In \bibinfo{booktitle}{\emph{Asia-Pacific
  Workshop on Systems (APSys)}}. \bibinfo{publisher}{ACM SIGOPS},
  \bibinfo{address}{Korea}.
\newblock


\bibitem[\protect\citeauthoryear{Gullasch, Bangerter, and Krenn}{Gullasch
  et~al\mbox{.}}{2011}]%
        {Gullasch_BK_11}
\bibfield{author}{\bibinfo{person}{David Gullasch}, \bibinfo{person}{Endre
  Bangerter}, {and} \bibinfo{person}{Stephan Krenn}.}
  \bibinfo{year}{2011}\natexlab{}.
\newblock \showarticletitle{Cache Games -- Bringing Access-Based Cache Attacks
  on {AES} to Practice}. In \bibinfo{booktitle}{\emph{Proceedings of the IEEE
  Symposium on Security and Privacy}}. \bibinfo{address}{Oakland, CA, US},
  \bibinfo{pages}{490--505}.
\newblock


\bibitem[\protect\citeauthoryear{Heiser}{Heiser}{2018}]%
        {Heiser_18}
\bibfield{author}{\bibinfo{person}{Gernot Heiser}.}
  \bibinfo{year}{2018}\natexlab{}.
\newblock \showarticletitle{For Safety's Sake: We Need a New Hardware-Software
  Contract!}
\newblock \bibinfo{journal}{\emph{IEEE Design and Test}}  \bibinfo{volume}{35}
  (\bibinfo{date}{March} \bibinfo{year}{2018}), \bibinfo{pages}{27--30}.
\newblock


\bibitem[\protect\citeauthoryear{Hill}{Hill}{2018}]%
        {Hill:ISA-blog}
\bibfield{author}{\bibinfo{person}{Mark~D. Hill}.}
  \bibinfo{year}{2018}\natexlab{}.
\newblock \showarticletitle{A Primer on the {Meltdown} \& {Spectre} Hardware
  Security Design Flaws and their Important Implications}.
\newblock \bibinfo{journal}{\emph{Computer Architecture Today}}
  (\bibinfo{date}{Feb.} \bibinfo{year}{2018}).
\newblock


\bibitem[\protect\citeauthoryear{Intel Corporation}{Intel Corporation}{2016}]%
        {Intel_64_IA-32:asdm2_325383}
Intel Corporation \bibinfo{year}{2016}\natexlab{}.
\newblock \bibinfo{booktitle}{\emph{{Intel} 64 and {IA-32} Architecture
  Software Developer's Manual Volume 2: Instruction Set Reference, A-Z}}.
\newblock Intel Corporation.
\newblock
\newblock
\shownote{\url{http://www.intel.com.au/content/dam/www/public/us/en/documents/manuals/64-ia-32-architectures-software-developer-instruction-set-reference-manual-325383.pdf}.}


\bibitem[\protect\citeauthoryear{Kessler and Hill}{Kessler and Hill}{1992}]%
        {Kessler_Hill_92}
\bibfield{author}{\bibinfo{person}{R.~E. Kessler} {and}
  \bibinfo{person}{Mark~D. Hill}.} \bibinfo{year}{1992}\natexlab{}.
\newblock \showarticletitle{Page placement algorithms for large real-indexed
  caches}.
\newblock \bibinfo{journal}{\emph{ACM Transactions on Computer Systems}}
  \bibinfo{volume}{10} (\bibinfo{year}{1992}), \bibinfo{pages}{338--359}.
\newblock


\bibitem[\protect\citeauthoryear{Klein, Andronick, Elphinstone, Murray, Sewell,
  Kolanski, and Heiser}{Klein et~al\mbox{.}}{2014}]%
        {Klein_AEMSKH_14}
\bibfield{author}{\bibinfo{person}{Gerwin Klein}, \bibinfo{person}{June
  Andronick}, \bibinfo{person}{Kevin Elphinstone}, \bibinfo{person}{Toby
  Murray}, \bibinfo{person}{Thomas Sewell}, \bibinfo{person}{Rafal Kolanski},
  {and} \bibinfo{person}{Gernot Heiser}.} \bibinfo{year}{2014}\natexlab{}.
\newblock \showarticletitle{Comprehensive Formal Verification of an {OS}
  Microkernel}.
\newblock \bibinfo{journal}{\emph{ACM Transactions on Computer Systems}}
  \bibinfo{volume}{32}, \bibinfo{number}{1} (\bibinfo{date}{Feb.}
  \bibinfo{year}{2014}), \bibinfo{pages}{2:1--2:70}.
\newblock


\bibitem[\protect\citeauthoryear{Klein, Murray, Gammie, Sewell, and
  Winwood}{Klein et~al\mbox{.}}{2011}]%
        {Klein_MGSW_11}
\bibfield{author}{\bibinfo{person}{Gerwin Klein}, \bibinfo{person}{Toby
  Murray}, \bibinfo{person}{Peter Gammie}, \bibinfo{person}{Thomas Sewell},
  {and} \bibinfo{person}{Simon Winwood}.} \bibinfo{year}{2011}\natexlab{}.
\newblock \showarticletitle{Provable Security: How feasible is it?}. In
  \bibinfo{booktitle}{\emph{Workshop on Hot Topics in Operating Systems}}.
  \bibinfo{publisher}{USENIX}, \bibinfo{address}{Napa, USA},
  \bibinfo{pages}{5}.
\newblock


\bibitem[\protect\citeauthoryear{Kocher, Horn, Fogh, Genkin, Gruss, Haas,
  Haburg, Lipp, Mangard, Prescher, Schwartz, and Yarom}{Kocher
  et~al\mbox{.}}{2019}]%
        {Kocher_HFGGHHLMPSY_19}
\bibfield{author}{\bibinfo{person}{Paul Kocher}, \bibinfo{person}{Jann Horn},
  \bibinfo{person}{Anders Fogh}, \bibinfo{person}{Daniel Genkin},
  \bibinfo{person}{Daniel Gruss}, \bibinfo{person}{Werner Haas},
  \bibinfo{person}{Mike Haburg}, \bibinfo{person}{Moritz Lipp},
  \bibinfo{person}{Stefan Mangard}, \bibinfo{person}{Thomas Prescher},
  \bibinfo{person}{Michael Schwartz}, {and} \bibinfo{person}{Yuval Yarom}.}
  \bibinfo{year}{2019}\natexlab{}.
\newblock \showarticletitle{Spectre Attacks: Exploiting Speculative Execution}.
  In \bibinfo{booktitle}{\emph{IEEE Symposium on Security and Privacy}}.
  \bibinfo{publisher}{IEEE}, \bibinfo{address}{San Francisco},
  \bibinfo{pages}{19--37}.
\newblock


\bibitem[\protect\citeauthoryear{Liedtke, H{\"a}rtig, and Hohmuth}{Liedtke
  et~al\mbox{.}}{1997}]%
        {Liedtke_HH_97}
\bibfield{author}{\bibinfo{person}{Jochen Liedtke}, \bibinfo{person}{Hermann
  H{\"a}rtig}, {and} \bibinfo{person}{Michael Hohmuth}.}
  \bibinfo{year}{1997}\natexlab{}.
\newblock \showarticletitle{{OS}-controlled cache predictability for real-time
  systems}. In \bibinfo{booktitle}{\emph{IEEE Real-Time and Embedded Technology
  and Applications Symposium (RTAS)}}. \bibinfo{publisher}{IEEE},
  \bibinfo{address}{Montreal, CA}, \bibinfo{pages}{213--223}.
\newblock


\bibitem[\protect\citeauthoryear{Lynch, Bray, and Flynn}{Lynch
  et~al\mbox{.}}{1992}]%
        {Lynch_BF_92}
\bibfield{author}{\bibinfo{person}{William~L. Lynch}, \bibinfo{person}{Brian~K.
  Bray}, {and} \bibinfo{person}{M.~J. Flynn}.} \bibinfo{year}{1992}\natexlab{}.
\newblock \showarticletitle{The effect of page allocation on caches}. In
  \bibinfo{booktitle}{\emph{ACM/IEE International Symposium on
  Microarchitecture}}. \bibinfo{pages}{222--225}.
\newblock


\bibitem[\protect\citeauthoryear{Murray, Matichuk, Brassil, Gammie, Bourke,
  Seefried, Lewis, Gao, and Klein}{Murray et~al\mbox{.}}{2013}]%
        {Murray_MBGBSLGK_13}
\bibfield{author}{\bibinfo{person}{Toby Murray}, \bibinfo{person}{Daniel
  Matichuk}, \bibinfo{person}{Matthew Brassil}, \bibinfo{person}{Peter Gammie},
  \bibinfo{person}{Timothy Bourke}, \bibinfo{person}{Sean Seefried},
  \bibinfo{person}{Corey Lewis}, \bibinfo{person}{Xin Gao}, {and}
  \bibinfo{person}{Gerwin Klein}.} \bibinfo{year}{2013}\natexlab{}.
\newblock \showarticletitle{{seL4}: from General Purpose to a Proof of
  Information Flow Enforcement}. In \bibinfo{booktitle}{\emph{IEEE Symposium on
  Security and Privacy}}. \bibinfo{address}{San Francisco, CA},
  \bibinfo{pages}{415--429}.
\newblock


\bibitem[\protect\citeauthoryear{Murray, Matichuk, Brassil, Gammie, and
  Klein}{Murray et~al\mbox{.}}{2012}]%
        {Murray_MBGK_12}
\bibfield{author}{\bibinfo{person}{Toby Murray}, \bibinfo{person}{Daniel
  Matichuk}, \bibinfo{person}{Matthew Brassil}, \bibinfo{person}{Peter Gammie},
  {and} \bibinfo{person}{Gerwin Klein}.} \bibinfo{year}{2012}\natexlab{}.
\newblock \showarticletitle{Noninterference for Operating System Kernels}. In
  \bibinfo{booktitle}{\emph{International Conference on Certified Programs and
  Proofs}}. \bibinfo{publisher}{Springer}, \bibinfo{address}{Kyoto, Japan},
  \bibinfo{pages}{126--142}.
\newblock


\bibitem[\protect\citeauthoryear{Osvik, Shamir, and Tromer}{Osvik
  et~al\mbox{.}}{2006}]%
        {Osvik_ST_06}
\bibfield{author}{\bibinfo{person}{Dag~Arne Osvik}, \bibinfo{person}{Adi
  Shamir}, {and} \bibinfo{person}{Eran Tromer}.}
  \bibinfo{year}{2006}\natexlab{}.
\newblock \showarticletitle{Cache Attacks and Countermeasures: The Case of
  {AES}}. In \bibinfo{booktitle}{\emph{Proceedings of the 2006 Crytographers'
  track at the RSA Conference on Topics in Cryptology}}.
\newblock


\bibitem[\protect\citeauthoryear{Percival}{Percival}{2005}]%
        {Percival_05}
\bibfield{author}{\bibinfo{person}{Colin Percival}.}
  \bibinfo{year}{2005}\natexlab{}.
\newblock \showarticletitle{Cache Missing for Fun and Profit}. In
  \bibinfo{booktitle}{\emph{BSDCon 2005}}. \bibinfo{address}{Ottawa, CA}.
\newblock


\bibitem[\protect\citeauthoryear{Syeda and Klein}{Syeda and Klein}{2018}]%
        {Syeda_Klein_18}
\bibfield{author}{\bibinfo{person}{Hira Syeda} {and} \bibinfo{person}{Gerwin
  Klein}.} \bibinfo{year}{2018}\natexlab{}.
\newblock \showarticletitle{Program Verification in the Presence of Cached
  Address Translation}. In \bibinfo{booktitle}{\emph{Interactive Theorem
  Proving, ITP}}, Vol.~\bibinfo{volume}{10895}. \bibinfo{publisher}{Springer's
  LNCS series}, \bibinfo{address}{Oxford, UK}, \bibinfo{pages}{542--559}.
\newblock


\bibitem[\protect\citeauthoryear{Wray}{Wray}{1991}]%
        {Wray_91}
\bibfield{author}{\bibinfo{person}{John~C. Wray}.}
  \bibinfo{year}{1991}\natexlab{}.
\newblock \showarticletitle{An analysis of covert timing channels}. In
  \bibinfo{booktitle}{\emph{Proceedings of the 1991 IEEE Computer Society
  Symposium on Research in Security and Privacy}}. \bibinfo{publisher}{IEEE},
  \bibinfo{address}{Oakland, CA, US}, \bibinfo{pages}{2--7}.
\newblock


\bibitem[\protect\citeauthoryear{Yarom and Falkner}{Yarom and Falkner}{2014}]%
        {Yarom_Falkner_14}
\bibfield{author}{\bibinfo{person}{Yuval Yarom} {and} \bibinfo{person}{Katrina
  Falkner}.} \bibinfo{year}{2014}\natexlab{}.
\newblock \showarticletitle{\textsc{Flush+Reload}: a High Resolution, Low
  Noise, {L3} Cache Side-Channel Attack}. In
  \bibinfo{booktitle}{\emph{Proceedings of the 23rd USENIX Security
  Symposium}}. \bibinfo{address}{San Diego, CA, US}, \bibinfo{pages}{719--732}.
\newblock


\end{thebibliography}
}
\end{document}